\documentclass[11pt,twoside]{article}             % Use with LaTeX2e
\usepackage{myosid}                         %
\usepackage{bbm,amsmath,amssymb,amsbsy,graphicx,subfigure}

\def\LHD{G. Adesso, A. Serafini, and F. Illuminati}
\def\RHD{Entanglement, Purity, and Information Entropies in Continuous Variable Systems}

\newcommand{\gr}[1]{\boldsymbol{#1}}

\newcommand{\abs}[1]{\left\vert#1\right\vert}
\renewcommand{\det}{{\rm Det}\,}
\newcommand{\eq}[1]{Eq.~(\ref{#1})}

\newcommand{\R}{\mathbbm R}
\newcommand{\N}{{\cal N}}

\title{Entanglement, Purity, and Information
Entropies \\in Continuous Variable Systems\footnote[2]{\ \ Presented
at the International Conference {\em ``Entanglement, Information \&
Noise''}, Krzy\.zowa, Poland, June 14--20, 2004.}}
\author{Gerardo Adesso, Alessio
Serafini, and Fabrizio Illuminati
\\[2ex]
{\footnotesize \sl Dipartimento di Fisica ``E. R. Caianiello'',
Universit\`a di Salerno, INFM UdR di Salerno, INFN Sezione di
Napoli, Gruppo Collegato di Salerno, Via S. Allende, 84081 Baronissi
(SA), Italy. e--mail: {gerardo@sa.infn.it}, {serale@sa.infn.it},
{illuminati@sa.infn.it}.}}
\begin{document}

\maketitle
\begin{abstract}
     Quantum entanglement of pure states of a bipartite system
     is defined as the amount of local or marginal
     ({\em i.e.~}referring to the subsystems) entropy. For mixed
     states this identification vanishes,
     since the global loss of information about the state
     makes it impossible to distinguish
     between quantum and classical correlations. Here we
     show how the joint knowledge of the global
     and marginal degrees of information of a quantum state,
     quantified by the purities or in general by
     information entropies, provides an accurate
     characterization of its entanglement.
     In particular, for Gaussian states of continuous variable
     systems, we classify the entanglement of two--mode states according to
their degree of total and partial mixedness, comparing the different
roles played by the purity and the generalized $p-$entropies in
quantifying the mixedness and bounding the entanglement. We prove
the existence of strict upper and lower bounds on the entanglement
and the existence of extremally (maximally and minimally) entangled
states at fixed global and marginal degrees of information. This
results allow for a powerful, operative method to measure
mixed-state entanglement without the full tomographic reconstruction
of the state. Finally, we briefly discuss the ongoing extension of
our analysis to the quantification of multipartite entanglement in
highly symmetric Gaussian states of arbitrary $1 \times N$-mode
partitions.
\end{abstract}

\section{Introduction}
According to Erwin Schr\"odinger, quantum entanglement is not
\emph{``one but rather the characteristic trait of quantum
mechanics, the one that enforces its entire departure from classical
lines of thought''} \cite{schr}. Entanglement has been widely
recognized as a fundamental aspect of quantum theory, stemming
directly from the superposition principle and quantum non
factorizability. Remarkably, it is now also acknowledged as a
fundamental physical resource, much on the same status as energy and
entropy, and as a key factor in the realization of information
processes otherwise impossible to implement on classical systems.
Thus the degree of entanglement and information are the crucial
features of a quantum state from the point of view of Quantum
Information Theory \cite{qinfobk}. Indeed, the search for proper
mathematical frameworks to quantify such features in generic (mixed)
quantum states cannot be yet considered accomplished. In view of
such considerations, it is clear that the full understanding of the
relationships between the quantum correlations contained in a
multipartite state and the global and local ({\em i.e.~}referring to
the reduced states of the subsystems) degrees of information of the
state, is of critical importance. In particular, it would represent
a relevant step towards the clarification of the nature of quantum
correlations and, possibly, of the distinction between quantum and
classical correlations of mixed states \cite{henvedral01}. The main
question we want to address in this work is:
\begin{quote}
 {What
can we say about the quantum correlations existing between the
subsystems of a quantum multipartite system in a mixed state, if we
know the degrees of information carried by the global and the
reduced states?}
\end{quote}
We can anticipate that the answer will be \emph{``almost
everything''} in the context of Gaussian states of continuous
variable systems. To this aim, we will start by briefly reviewing in
Sec.~\ref{secgauss} the properties of Gaussian states in
infinite--dimensional Hilbert spaces, and the concepts of
information and entanglement. In Sec.~\ref{secduemodi} we will show,
step by step, how the entanglement of two--mode Gaussian states can
be accurately characterized by the knowledge of global and marginal
degrees of information, quantified by the purities, or by the
generalized entropies of the global state and of the reduced states
of the subsystems. In Sec.~\ref{secmulti} we will give a brief
sketch of the generalization of our methods to the quantification of
multipartite entanglement in multimode Gaussian states under
symmetry. Finally, in Sec.~\ref{secconcl} we will summarize our
results and discuss future perspectives.

\section{Gaussian states: general properties}\label{secgauss}

We consider a continuous variable (CV) system consisting of $N$
canonical bosonic modes, associated to an infinite-dimensional
Hilbert space and described by the vector $\hat{X}=\{\hat x_1,\hat
p_1,\ldots,\hat x_N,\hat p_N\}$ of the field quadrature
(``position'' and ``momentum'') operators. The quadrature phase
operators are connected to the annihilation $\hat a_i$ and creation
$\hat a^{\dag}_i$ operators of each mode, by the relations $\hat
x_{i}=(\hat a_{i}+\hat a^{\dag}_{i})$ and $\hat p_{i}=(\hat
a_{i}-\hat a^{\dag}_{i})/i$. The canonical commutation relations for
the $\hat X_i$'s can be expressed in matrix form: $[\hat X_{i},\hat
X_j]=2i\Omega_{ij}$, with the symplectic form
$\Omega=\oplus_{i=1}^{n}\omega$ and $\omega=\delta_{ij-1}-
\delta_{ij+1},\, i,j=1,2$.

Quantum states of paramount importance in CV systems are the
so-called Gaussian states, {\em i.e.}~states with Gaussian
characteristic functions and quasi--probability distributions
\cite{cvbook}. The interest in this special class of states
(important examples are vacua, coherent, squeezed and thermal states
of the electromagnetic field) stems from the feasibility to produce
and control them with linear optics, and from the increasing number
of efficient proposals and successful experimental implementations
of CV quantum information and communication processes involving
multimode Gaussian states (see \cite{review} for a recent review).
By definition, a Gaussian state  is completely characterized by
first and second moments of the canonical operators. When addressing
physical properties invariant under local unitary transformations,
such as the mixedness and the entanglement, one can neglect first
moments and completely characterize Gaussian states by the $2N\times
2N$ real covariance matrix (CM) $\gr{\sigma}$, whose entries are
$\sigma_{ij}=1/2\langle\{\hat{X}_i,\hat{X}_j\}\rangle
-\langle\hat{X}_i\rangle\langle\hat{X}_j\rangle$. Throughout the
paper, $\gr{\sigma}$ will be used indifferently to indicate the CM
of a Gaussian state or the state itself. A real, symmetric matrix
$\gr{\sigma}$ must fulfill the Robertson-Schr\"odinger uncertainty
relation \cite{simon87}
\begin{equation}\label{bonafide}
\gr{\sigma}+i\Omega \geq 0\,,
\end{equation}
to be a {\em bona fide} CM of a physical state. Symplectic
operations ({\em i.e.}~belonging to the group $Sp_{(2N,\R)}= \{S\in
SL(2N,\R)\,:\,S^T\Omega S=\Omega\}$) acting by congruence on CMs in
phase space, amount to unitary operations on density matrices in
Hilbert space. In phase space, any $N$-mode Gaussian state can be
transformed by symplectic operations in its Williamson diagonal form
 $\gr\nu$ \cite{williamson36}, such that $\gr{\sigma}= S^T \gr{\nu} S$,
with $\gr{\nu}=\,{\rm diag}\,\{\nu_1,\nu_1,\ldots\nu_N,\nu_N\}$. The
set $\Sigma=\{\nu_i\}$ constitutes the symplectic spectrum of
$\gr{\sigma}$ and its elements must fulfill the conditions $\nu_i\ge
1$, following from \eq{bonafide} and ensuring positivity of the
density matrix associated to $\gr{\sigma}$. We remark that the full
saturation of the uncertainty principle can only be achieved by pure
$N$-mode Gaussian states, for which $\nu_i=1\,\,\forall i=1,\ldots,
N$. Instead, mixed states such that $\nu_{i\le k}=1$ and
$\nu_{i>k}>1$, with $1\le k\le N$, only partially saturate the
uncertainty principle, with partial saturation becoming weaker with
decreasing $k$. The symplectic eigenvalues $\nu_i$ can be computed
as the orthogonal eigenvalues of the matrix $|i\Omega\gr{\sigma}|$,
so they are determined by $N$ symplectic invariants associated to
the characteristic polynomial of such a matrix. Two global
invariants which will be useful are the determinant $\,{\rm
Det}\,\gr{\sigma}=\prod_{i}\nu_i^2$ and the \emph{seralian}
\cite{serafozzi} $\Delta=\sum_i \nu_i^2$, which is the sum of the
determinants of all the $2\times 2$ submatrices of $\gr{\sigma}$
related to each mode.

The degree of information about the preparation of a quantum state
$\varrho$ can be characterized by its \emph{purity} $\mu\equiv\,{\rm
Tr}\,\varrho^2$. For a Gaussian state with CM $\gr{\sigma}$ one has
simply $\mu=1/\sqrt{\,{\rm Det}\,\gr{\sigma}}$ \cite{paris}. In
general, the mixedness, or lack of information about the preparation
of the state, can be quantified by generalized entropic measures,
such as the Bastiaans--Tsallis entropies \cite{bastiaans,tsallis}
$S_{p} \equiv (1-\,{\rm Tr}\,\varrho^p)/(p-1)$, which reduce to the
linear entropy $S_L=1-\mu$ for $p=2$, and the R\'{e}nyi entropies
\cite{renyi} $S_{p}^{R} \equiv (\log \, {\rm Tr} \,
\varrho^{p})/(1-p)$. Both entropic families are parametrized by $p >
1$ and it can be easily shown that
$\lim_{p\rightarrow1+}S_{p}=\lim_{p\rightarrow1+}S_{p}^{R}= -\,{\rm
Tr}\,(\varrho\log\varrho) \equiv S_{V}$, so that also the
Shannon-von Neumann entropy $S_{V}$ can be defined in terms of
generalized entropies. The quantity $S_{V}$ is additive on tensor
product states and provides a further convenient measure of
mixedness of the quantum state $\varrho$.

As for the entanglement, we recall that positivity of the CM's
partial transpose (PPT) \cite{ppt} is a necessary and sufficient
condition of separability for $(N+1)$-mode Gaussian states with
respect to $1\times N$-mode partitions \cite{simon00}. In phase
space, partial transposition amounts to a mirror reflection of one
quadrature associated to the single-mode partition. If
$\{\tilde{\nu}_i\}$ is the symplectic spectrum of the partially
transposed CM $\tilde{\gr{\sigma}}$, then a $(N+1)$-mode Gaussian
state with CM $\gr{\sigma}$ is separable if and only if
$\tilde{\nu}_i\ge 1$ $\forall\, i$. A convenient measure of CV
entanglement is the \emph{logarithmic negativity} \cite{vidwer}
$E_{\N}\equiv \log\|\tilde{\varrho}\|_{1}$, $\| \cdot \|_1$ denoting
the trace norm, which constitutes an upper bound to the {\em
distillable entanglement} of the quantum state $\varrho$. It can be
readily computed in terms of the symplectic spectrum $\tilde{\nu}_i$
of $\tilde{\gr{\sigma}}$, yielding \cite{asimulti}
\begin{equation}\label{ensy}
E_{\N}=\left\{%
\begin{array}{cc}
    0, & \tilde{\nu}_i \ge 1\;\forall\,i\,; \\
    -\sum_{i :
\tilde{\nu}_i<1}\log\tilde{\nu}_i\,, & \hbox{else .} \\
\end{array}%
\right.
\end{equation}
The logarithmic negativity quantifies the extent to which the PPT
condition $\tilde{\nu}_i\ge 1$ is violated.

\section{Characterizing two--mode entanglement by information measures}
\label{secduemodi}
\subsection{Parametrization of Gaussian states with symplectic invariants}
Two--mode Gaussian states represent the prototypical quantum states
of CV systems. Their CM can be written is the following block form
\begin{equation}
\boldsymbol{\sigma}\equiv\left(\begin{array}{cc}
\boldsymbol{\alpha}&\boldsymbol{\gamma}\\
\boldsymbol{\gamma}^{T}&\boldsymbol{\beta}
\end{array}\right)\, , \label{espre}
\end{equation}
where the three $2\times 2$ matrices $\boldsymbol{\alpha}$,
$\boldsymbol{\beta}$, $\boldsymbol{\gamma}$ are, respectively, the
CMs of the two reduced modes and the correlation matrix between
them. It is well known that for any two--mode CM
$\boldsymbol{\sigma}$ there exists a local symplectic operation
$S_{l}=S_{1}\oplus S_{2}$ which takes $\boldsymbol{\sigma}$ to the
so called standard form $\boldsymbol{\sigma}_{sf}$ \cite{duan00}
\begin{equation}
S_{l}^{T}\boldsymbol{\sigma}S_{l}=\boldsymbol{\sigma}_{sf} \equiv
\left(\begin{array}{cccc}
a&0&c_{+}&0\\
0&a&0&c_{-}\\
c_{+}&0&b&0\\
0&c_{-}&0&b
\end{array}\right)\; . \label{stform}
\end{equation}
States whose standard form fulfills $a=b$ are said to be symmetric.
Let us recall that any pure state is symmetric and fulfills
$c_{+}=-c_{-}=\sqrt{a^2-1}$.  The uncertainty principle
Ineq.~(\ref{bonafide}) can be recast as a constraint on the
$Sp_{(4,{\mathbb R})}$ invariants ${\rm Det}\gr{\sigma}$ and
$\Delta(\gr{\sigma})={\rm Det}\boldsymbol{\alpha}+\,{\rm
Det}\boldsymbol{\beta}+2 \,{\rm Det}\boldsymbol{\gamma}$, yielding
%\begin{equation}\label{sympheis}
$\Delta(\gr{\sigma})\le1+\,{\rm Det}\boldsymbol{\sigma}$.
%\end{equation}
The symplectic eigenvalues of a two--mode Gaussian state will be
named $\nu_{-}$ and $\nu_{+}$, with $\nu_- \le \nu_+$ in general. A
simple expression for the $\nu_{\mp}$ can be found in terms of the
two $Sp_{(4,\mathbb{R})}$ invariants \cite{serafozzi}
\begin{equation}
2{\nu}_{\mp}^2=\Delta(\gr{\sigma})\mp\sqrt{\Delta(\gr{\sigma})^2
-4\,{\rm Det}\,\gr{\sigma}} \, . \label{sympeig}
\end{equation}

The standard form covariances $a$, $b$, $c_{+}$, and $c_{-}$ can be
determined in terms of the two local symplectic invariants
\begin{equation}\label{mu12}
\mu_1 = (\det\gr\alpha)^{-1/2} = 1/a\,,\quad \mu_2 =
(\det\gr\beta)^{-1/2} = 1/b\,,
\end{equation}
which are the marginal purities of the reduced single--mode states,
and of the two global symplectic invariants
\begin{equation}\label{globinv}
\mu = (\det\gr\sigma)^{-1/2} =
[(ab-c_{+}^2)(ab-c_{-}^2)]^{-1/2}\,,\quad \Delta =
a^2+b^2+2c_+c_-\,,
\end{equation}
which are the global purity and the seralian, respectively.
Eqs.~(\ref{mu12}-\ref{globinv}) can be inverted to provide the
following physical parametrization of two--mode states in terms of
the four independent parameters $\mu_1,\,\mu_2,\,\mu$, and $\Delta$
\cite{asipra04}:
\begin{eqnarray}
% \nonumber to remove numbering (before each equation)
  a  \,\,=\,\, \frac{1}{\mu_1}\,, \quad b &=& \frac{1}{\mu_2}\,, \quad
c_{\pm}\,\,=\,\,\frac{\sqrt{\mu_1 \mu_2}}4 \, \big( \epsilon_- \pm
\epsilon_+ \big)\,, \label{gabc} \\
{\rm with}\quad \epsilon_\mp &\equiv& \sqrt{\left[ \Delta -
\frac{(\mu_1 \mp \mu_2)^2}{\mu_1^2 \mu_2^2}\right]^2-\frac{4}{\mu^2}
}\, . \nonumber
\end{eqnarray}

The uncertainty principle and the existence
of the radicals appearing in \eq{gabc} impose the following
constraints on the four invariants in order to describe a physical
state
\begin{eqnarray}
  0 &\le& \mu_{1,2} \,\,\le\,\, 1\,, \label{consmu12}\\
  \mu_1 \mu_2 &\le& \mu \,\,\le\,\,
  \frac{\mu_1 \mu_2}{\mu_1 \mu_2 + \abs{\mu_1-\mu_2}}\,, \label{consmu} \\
  \frac{2}{\mu} + \frac{(\mu_1 - \mu_2)^2}{\mu_1^2 \mu_2^2}
  &\le& \Delta  \,\,\le\,\,  1+\frac{1}{\mu^2}  \, . \label{deltabnd}
\end{eqnarray}
The physical meaning of these constraints, and the role of the
extremal states ({\em i.e.~}states whose invariants saturate the
upper or lower bounds of Eqs.~(\ref{consmu}-\ref{deltabnd})) in
relation to the entanglement, will be carefully investigated in the
next subsections.

In terms of symplectic invariants, partial transposition corresponds
to flipping the sign of ${\rm Det}\,\gr{\gamma}$, so that $\Delta$
turns into $\tilde{\Delta}=\Delta-4\,{\rm Det}\,\gr{\gamma} =
-\Delta + 2/\mu_1^2 + 2/\mu_2^2$. The symplectic eigenvalues of the
CM $\gr{\sigma}$ and of its partial transpose
$\tilde{\gr{\sigma}}$ are promptly determined in terms of symplectic
invariants
\begin{equation}
  2\nu_{\mp}^2 = \Delta\mp\sqrt{\Delta^2
-{\frac{4}{\mu^2}}}\,, \quad
  2\tilde{\nu}_{\mp}^2 = \tilde{\Delta}\mp\sqrt{\tilde{\Delta}^2
-{\frac{4}{\mu^2}}}\,. \label{n1}
\end{equation}
The PPT criterion yields a state $\gr{\sigma}$ separable if and only
if
%\begin{equation}
$\tilde{\nu}_{-}\ge 1$.
%\label{lowest}
%\end{equation}
A {\it bona fide} measure of entanglement for two--mode Gaussian
states should thus be a monotonically decreasing function of
$\tilde{\nu}_{-}$ \cite{notepad}, quantifying the violation of
the previous inequality. A computable entanglement monotone for generic
two-mode Gaussian states is provided by the logarithmic negativity
\eq{ensy}
\begin{equation}\label{en}
E_{\N}=\max\{0,-\log\,\tilde{\nu}_{-}\}\,.
\end{equation}
In the special instance of symmetric Gaussian states, the
\emph{entanglement of formation} \cite{entfor} is also computable
\cite{giedke03} but, being again a decreasing function of
$\tilde\nu_-$, it provides the same characterization of entanglement
and is thus fully equivalent to the logarithmic negativity in this
subcase.

\subsection{Entanglement vs Information (I) -- Maximal entanglement at fixed global purity}

The first step towards giving an answer to our original
question is to investigate the properties of extremally
entangled states at a given degree of global information. Let us
mention that, for two--qubit systems, the existence of maximally
entangled states at fixed mixedness (MEMS) was first found numerically
by Ishizaka and Hiroshima \cite{hishizaka00}. The discovery of such
states spurred several theoretical works \cite{vers01}, aimed at
exploring the relations between different measures of entanglement
and mixedness \cite{wei03} (strictly related to the questions of the
ordering of these different measures \cite{order}, and to the volume
of the set of mixed entangled states \cite{volume}).

Unfortunately, it is easy to show that a similar analysis in the CV
scenario is meaningless. Indeed, for any fixed, finite global purity
$\mu$ there exist infinitely many Gaussian states which are
infinitely entangled. As an example, we can consider the class of
(nonsymmetric) two--mode squeezed thermal states. Let
$S_{r}=\exp(\frac12 r a_{1}a_{2}-\frac12 r
a_{1}^{\dag}a_{2}^{\dag})$ be the two mode squeezing operator with
real squeezing parameter $r\ge0$, and let
$\varrho^{_\otimes}_{\nu_i}$ be a tensor product of thermal states
with CM ${\gr\nu}_{\nu_{\mp}}= {\mathbbm 1}_{2}\nu_- \oplus
{\mathbbm 1}_{2}\nu_{+}$, where $\nu_{\mp}$ is, as usual, the
symplectic spectrum of the state. Then, a nonsymmetric two-mode
squeezed thermal state $\xi_{\nu_{i},r}$ is defined as
${\xi}_{\nu_{i},r}=S_{r}\varrho^{_\otimes}_{\nu_i}S_{r}^{\dag}$,
corresponding to a standard form with
\begin{eqnarray}
a&=&{\nu_-}\cosh^{2}r+{\nu_{+}}\sinh^{2}r \; ,\quad
b\,\,=\,\,{\nu_{-}}\sinh^{2}r+{\nu_{+}}\cosh^{2}r \; ,\label{2mst}\\
c_{\pm}&=&\pm\frac{\nu_-+\nu_+}{2}\sinh2r \;. \nonumber
\end{eqnarray}
For simplicity we can consider the symmetric instance
($\nu_-=\nu_+=1/\sqrt\mu$) and compute the logarithmic negativity
\eq{en}, which takes the expression $E_\N(r,\mu) = -(1/2)\log[{\rm
e}^{-4r}/\mu]$. Notice how the completely mixed state ($\mu
\rightarrow 0$) is always separable while, for any $\mu > 0$, we can
freely increase the squeezing $r$ to obtain Gaussian states with
arbitrarily large entanglement. For fixed squeezing, as naturally
expected, the entanglement decreases with decreasing degree of
purity of the state, analogously to what happens in
discrete--variable MEMS \cite{wei03}.

\subsection{Entanglement vs Information (II) -- Maximal entanglement at fixed local purities}

The next step in the analysis is the unveiling of the relation
between the entanglement of a Gaussian state of CV systems and the
degrees of information related to the subsystems. Maximally
entangled states for given marginal mixednesses (MEMMS) have been
recently introduced and analyzed in detail in the context of qubit
systems by Adesso {\em et al.} \cite{adesso03}. The MEMMS provide a
suitable generalization of pure states, in which the entanglement is
completely quantified by the marginal degrees of mixedness.

 For two--mode Gaussian states, it follows from the expression \eq{n1} of
$\tilde\nu_-$ that, for fixed marginal purities $\mu_{1,2}$ and
seralian $\Delta$, the logarithmic negativity is strictly increasing
with increasing $\mu$. By imposing the saturation of the upper bound
of \eq{consmu}, $\mu=\mu^{\max}(\mu_{1,2})\equiv(\mu_1 \mu_2)/(\mu_1
\mu_2 + \abs{\mu_1-\mu_2})$, we determine the most pure states for
fixed marginals; moreover, choosing $\mu=\mu^{\max}(\mu_{1,2})$
immediately implies that the upper and the lower bounds on $\Delta$
of \eq{deltabnd} coincide and $\Delta$ is uniquely determined in
terms of $\mu_{1,2}$. This means that the two--mode states with
maximal purity for fixed marginals are indeed the Gaussian maximally
entangled states for fixed marginal mixednesses (GMEMMS). They can
be seen as the CV analogues of the MEMMS.
\begin{figure*}[t!]
\centering
 \subfigure[] {\includegraphics[width=6.2cm]{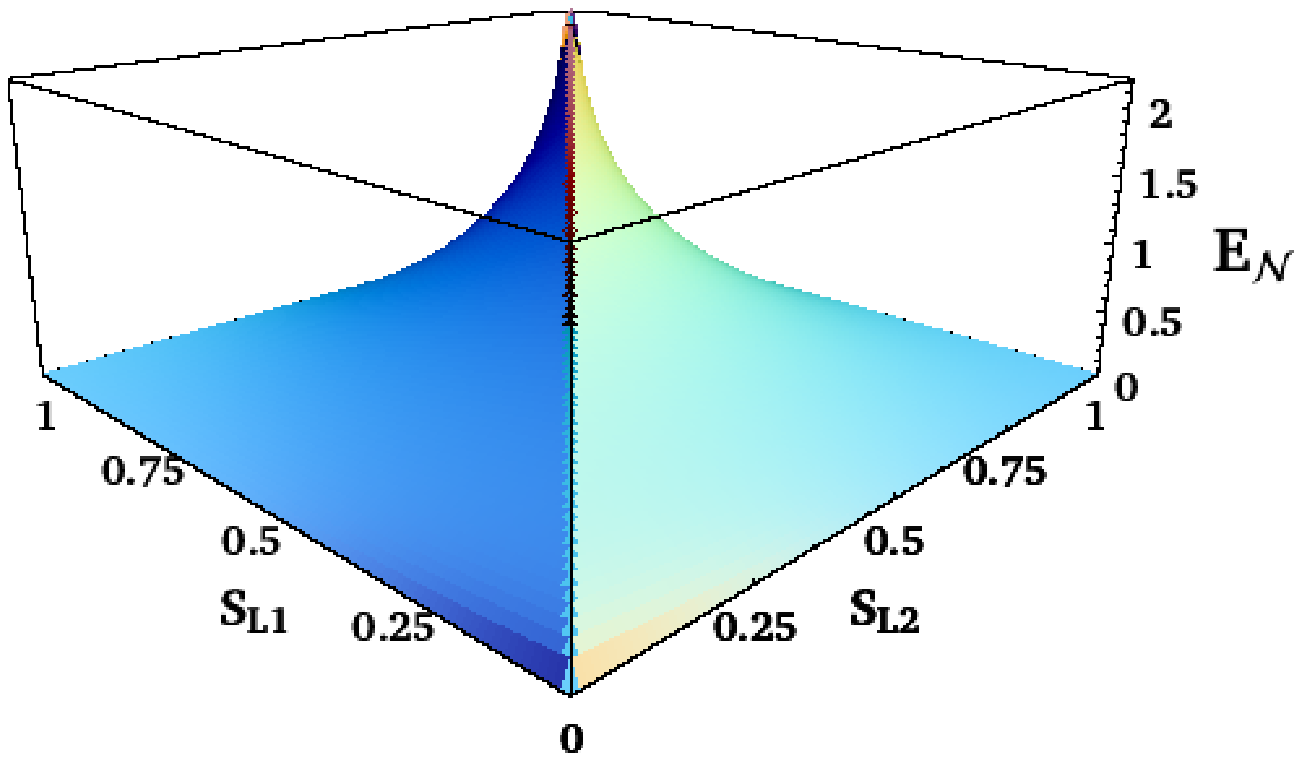}} \hspace{.5cm}
\subfigure[] {\includegraphics[width=6.2cm]{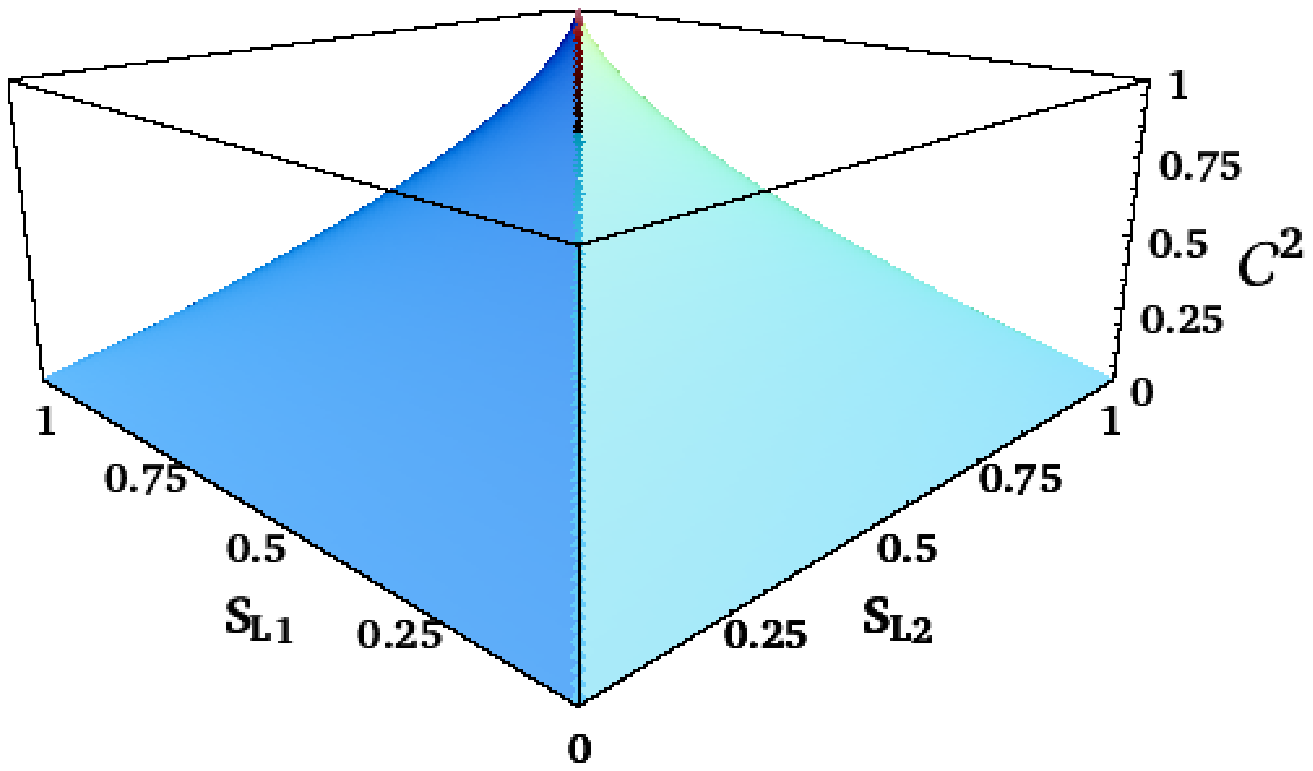}}
\caption{\footnotesize Plot of the maximal entanglement achievable
by quantum systems with given
 marginal linear entropies: (a) logarithmic negativity of continuous
 variable GMEMMS, which saturate the upper bound of inequality (\ref{consmu});
 (b) tangle of two--qubit MEMMS, introduced in Ref.~\cite{adesso03}.}
 \label{gmemms}
\end{figure*}
In Fig.~\ref{gmemms} the
logarithmic negativity of GMEMMS is plotted (a) as a function of the
marginal linear entropies $S_{L{1,2}} \equiv 1- \mu_{1,2}$, in
comparison (b) with the behaviour of the tangle (an entanglement
monotone equivalent to the entanglement of formation for two qubits
\cite{wootters}) as a function of $S_{L{1,2}}$ for discrete variable
MEMMS. Notice, as a common feature, how the maximal entanglement
achievable by quantum mixed states rapidly increases with increasing
marginal mixednesses (like in the pure--state instance) and
decreases with increasing difference of the marginals. This is
natural, because the presence of quantum correlations between the
subsystems implies that they should possess similar amounts of
quantum information. Let us finally mention that the ``minimally''
entangled states for fixed marginals, which saturate the lower bound
of \eq{consmu} ($\mu=\mu_1 \mu_2$), are just the tensor product
states, {\em i.e.~}states without any (quantum or classical)
correlations between the subsystems.

\subsection{Entanglement vs Information (III) --
Extremal entanglement at fixed global and local
purities}\label{secprl}

What we have shown so far, by simple analytical bounds, is a general
trend of increasing entanglement with increasing global purity, and
with decreasing marginal purities and difference between them. We
now wish to exploit the joint information about global and marginal
degrees of purity to achieve a significative characterization of
entanglement, both qualitatively and quantitatively. Let us first
investigate the role played by the seralian $\Delta$ in the
characterization of the properties of Gaussian states. To this aim,
we analyse the dependence of the eigenvalue $\tilde{\nu}_-$ on
$\Delta$, for fixed $\mu_{1,2}$ and $\mu$:
\begin{equation}
% \nonumber to remove numbering (before each equation)
  \left. \frac{\partial\ \tilde{\nu}^2_{-}}{\partial\ \Delta}
  \right|_{\mu_1,\,\mu_2,\,\mu} \,
 = \; \frac12 \left(
\frac{\tilde{\Delta}}{\sqrt{\tilde{\Delta}^2 -{\frac{1}{4 \mu^2}}}}
-1 \right) \,
> 0 \, .
\label{derivata}\end{equation}
The smallest symplectic eigenvalue of
the partially transposed state is strictly monotone in $\Delta$.
Therefore the entanglement of a generic Gaussian state $\gr{\sigma}$
with given global purity $\mu$ and marginal purities $\mu_{1,2}$, strictly
increases with decreasing $\Delta$. The seralian $\Delta$ is thus
endowed with a direct physical interpretation: at given global and
marginal purities, it determines the amount of entanglement of the
state. Moreover, due to inequality (\ref{deltabnd}), $\Delta$ is constrained
both by lower and upper bounds; therefore, not only maximally but also
\emph{minimally} entangled Gaussian states exist. This fact admirably
elucidates the relation between quantum correlations and information
in two--mode Gaussian states: the entanglement of such states is
tightly bound by the amount of global and marginal purities, with
only one remaining degree of freedom related to the invariant
$\Delta$ \cite{adeser04}.

We now aim to characterize \emph{extremally} (maximally and
minimally) entangled Gaussian states for fixed global and marginal
purities. Let us first consider the states saturating the lower
bound in Eq.~(\ref{deltabnd}), which entails \emph{maximal}
entanglement and defines the class of Gaussian most entangled states
for fixed global and local purities (GMEMS). It is easily seen that
such states belong to the class of asymmetric two--mode squeezed
thermal states \eq{2mst}, with squeezing parameter and symplectic
spectrum
\begin{eqnarray}\label{2sqp}
\tanh 2r &=& 2(\mu_1\mu_2-\mu_1^2\mu_2^2/\mu)^{1/2}/(\mu_1+\mu_2)\,
, \\
\nu_{\mp}^2 &=&
\frac{1}{\mu}+\frac{(\mu_1-\mu_2)^2}{2\mu_1^2\mu_2^2}\mp
\frac{|\mu_1-\mu_2|}{2\mu_1\mu_2}\sqrt{\frac{(\mu_1-\mu_2)^2}{\mu_1^2\mu_2^2}
+\frac{4}{\mu}}\, .
\end{eqnarray}
Nonsymmetric two--mode thermal squeezed states turn out to be
\emph{separable} in the range
\begin{equation}
\label{gnsmsep} \mu \le \frac{\mu_1 \mu_2}{\mu_1 + \mu_2 - \mu_1
\mu_2} \, .
\end{equation}
As a consequence, all Gaussian states whose purities fall in the
\emph{separable region} defined by inequality (\ref{gnsmsep}) are
not entangled.

\begin{figure}[t!]
\hspace*{-.5cm}
\begin{minipage}[b!]{7.5cm}
\includegraphics[width=7.2cm]{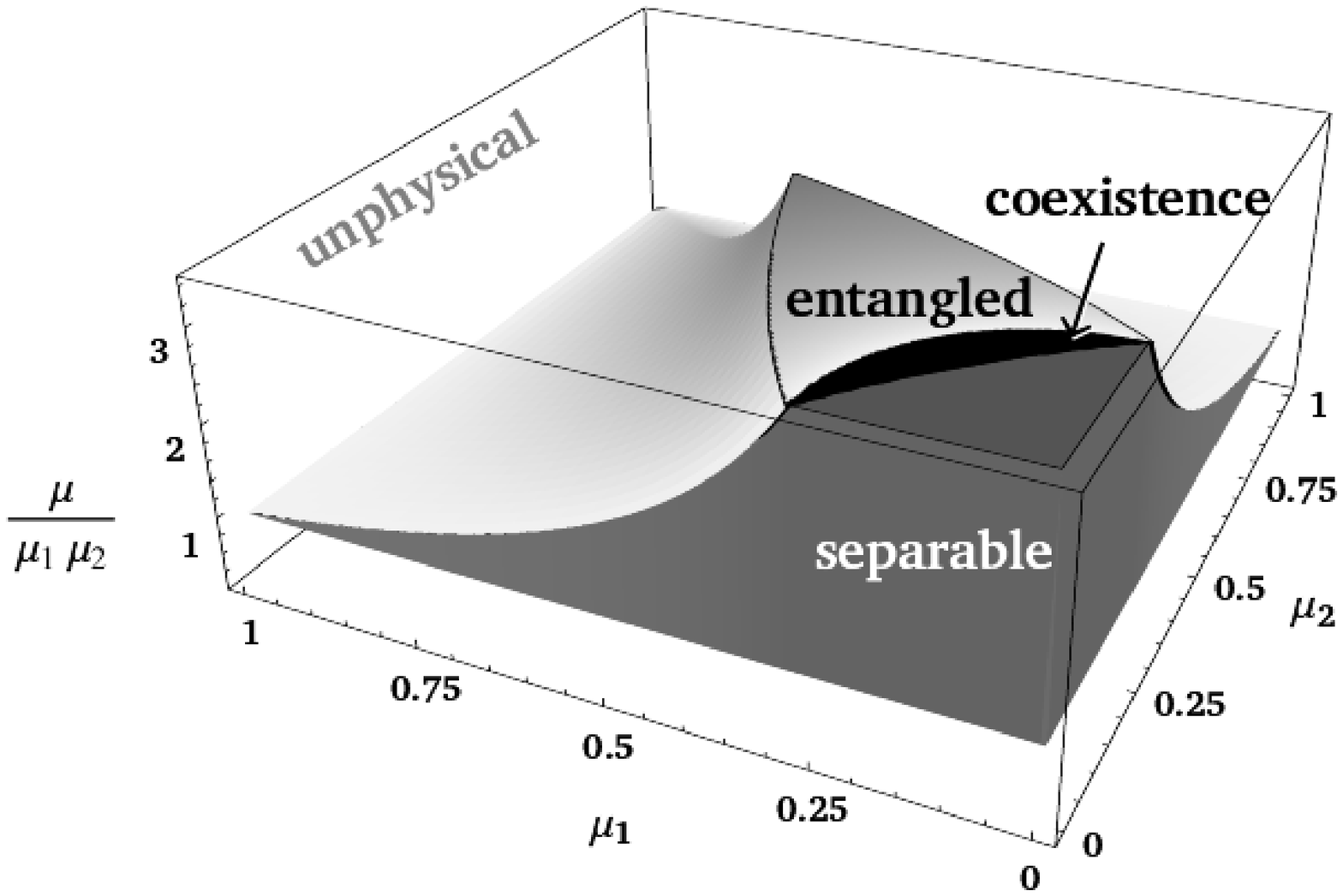}
\end{minipage}
\begin{minipage}[b!]{7cm}
\centering {\scriptsize
  \begin{tabular}{|c|c|}
  % after \\: \hline or \cline{col1-col2} \cline{col3-col4} ...
  \hline
  \textsc{Degrees of Purity} & \textsc{Regions}\\ \hline \hline
  $\mu<\mu_1 \mu_2$ & unphysical\\
  \hline
  $\mu_1 \mu_2 \; \le \; \mu \; \le \; \frac{\mu_1 \mu_2}{\mu_1 + \mu_2 - \mu_1 \mu_2}$ &
  \emph{separable}
  \\ \hline
  $\frac{\mu_1 \mu_2}{\mu_1 + \mu_2 - \mu_1 \mu_2} < \mu \le
  \frac{\mu_1 \mu_2}{\sqrt{\mu_1^2 + \mu_2^2 - \mu_1^2 \mu_2^2}}$ &
  \emph{coexistence}\\
   \hline
  $\frac{\mu_1 \mu_2}{\sqrt{\mu_1^2 + \mu_2^2 - \mu_1^2 \mu_2^2}} <
  \mu \le \frac{\mu_1 \mu_2}{\mu_1 \mu_2 + \abs{\mu_1-\mu_2}}$ &
  \emph{entangled}
  \\ \hline
  $\mu > \frac{\mu_1 \mu_2}{\mu_1 \mu_2 + \abs{\mu_1-\mu_2}}$ &
  unphysical
  \\ \hline
\end{tabular}}\\
\vspace*{.5cm} {\footnotesize Table I}
\end{minipage}
 \caption{\footnotesize
  Summary of entanglement properties of two--mode
  Gaussian states in the space of marginal purities $\mu_{1,2}$
  ($x$- and $y$-axes) and global purity $\mu$ (on the $z$-axis
  we plot the ratio $\mu/\mu_1\mu_2$ to gain a better graphical
  distinction between the various regions). In this space, all physical
  states lay between the horizontal plane $z=1$ representing product states,
  and the upper limiting surface representing GMEMMS (see Fig.~\ref{gmemms}a). Separable and entangled
  states are well distinguished except for a narrow coexistence region (depicted
  in black). The first ones fill the region depicted in dark grey,
  while in the area containing only entangled states the average logarithmic negativity
  Eq.~(\ref{average}) grows from white to
  medium grey.
  The mathematical relations defining the boundaries between all these regions
  are collected in Table I.
  The 3D envelope is cut at $z=3.5$.}
  \label{fig2D}
\end{figure}

We next consider the states that saturate the upper bound in
Eq.~(\ref{deltabnd}). They determine the class of Gaussian least
entangled states for given global and local purities (GLEMS) and,
outside the separable region (where every Gaussian state can be
considered a GLEMS having zero entanglement), they fulfill
%\begin{equation} \label{glm}
$\Delta = 1 + 1/{\mu}^{2}$.
%\end{equation}
This relation implies that the symplectic spectrum of these states
takes the form $\nu_-=1$, $\nu_+=1/\mu$. We thus find that GLEMS are
mixed Gaussian states of partial minimum uncertainty, so in some sense they
are the most classical ones and this is consistent with their
property of having minimal entanglement. According to the PPT
criterion, GLEMS are separable only if $\mu \le \mu_1 \mu_2 /
\sqrt{\mu_1^2 + \mu_2^2 - \mu_1^2 \mu_2^2}$. Therefore, in the range
\begin{equation}
\label{gnslsep} \frac{\mu_1 \mu_2}{\mu_1 + \mu_2 - \mu_1 \mu_2} <
\mu \le \frac{\mu_1 \mu_2}{\sqrt{\mu_1^2 + \mu_2^2 - \mu_1^2
\mu_2^2}}
\end{equation}
both separable and entangled states can be found. Instead, the
region
\begin{equation} \mu>\frac{\mu_1 \mu_2}{\sqrt{\mu_1^2 +
\mu_2^2 - \mu_1^2 \mu_2^2}} \label{sufent}
\end{equation}
can only accomodate \emph{entangled} states. The very narrow region
defined by inequality (\ref{gnslsep}) is thus the only \emph{region
of coexistence} of both entangled and separable Gaussian mixed
states. The discrimination of the different zones provides strong
necessary or sufficient conditions for the entanglement in terms of
the degrees of information, and allows to classify the separability of
all two-mode Gaussian states according to their global and marginal
purities, as shown in Fig.~\ref{fig2D}.

Knowledge of $\mu_{1,2}$ and $\mu$ thus accurately \emph{qualifies}
the entanglement of Gaussian states: as we will now show, quantitative
knowledge of the local and
global purities provides a reliable \emph{quantification} of
entanglement as well. Outside the separable region (see Table I in
Fig.~\ref{fig2D}), GMEMS attain maximum logarithmic negativity
$E_{{\N}max}(\mu_{1,2},\mu)$, while, in the entangled region, GLEMS
acquire minimum logarithmic negativity $E_{{\N}min}(\mu_{1,2},\mu)$,
where
\begin{eqnarray}
% \nonumber to remove numbering (before each equation)
\!\!\!E_{{\N}max} \!&\!\!\!=\!\!\!&\! -\! \frac12\log\!
\left[-\frac{1}{\mu}
  + \left(\frac{\mu_1+\mu_2}{2\mu_1^2 \mu_2^2} \right)\left(\mu_1+\mu_2 -
  \sqrt{(\mu_1+\mu_2)^2-\frac{4 \mu_1^2 \mu_2^2}{\mu}} \right) \right]\!\!,
\label{enmax} \\
\!\!\!E_{{\N}min} \!&\!\!\!=\!\!\!&\! -\! \frac12 \log\!
\left[\frac{1}{\mu_1^2}+\frac{1}{\mu_2^2}-\frac{1}{2\mu^2} -
  \frac12
- \sqrt{\left( \frac{1}{\mu_1^2}+\frac{1}{\mu_2^2}- \frac{1}{2\mu^2}
- \frac12 \right)^2 - \frac{1}{\mu^2}} \right]\!\!. \label{enmin}
\end{eqnarray}
Knowledge of the full CM, {\em i.e.~}including the symplectic
invariant $\Delta$ or all the cross-correlations, would allow for an
exact quantification of the entanglement. However, we will now show
that an estimate based only on the knowledge of the experimentally
measurable global and marginal purities turns out to be quite
accurate. We will quantify the entanglement of Gaussian states with
given global and marginal purities by the \emph{``average
logarithmic negativity''} $\bar{E}_{\N}$ \cite{adeser04}
\begin{equation}
\bar{E}_{\N} (\mu_{1,2},\mu) \equiv
\frac{E_{{\N}max}(\mu_{1,2},\mu)+E_{{\N}min}(\mu_{1,2},\mu)}{2} \; .
\label{average}
\end{equation}
We can then also define the relative error $\delta \bar{E}_{\N}$ on
$\bar{E}_{\N}$ as
\begin{equation}\label{deltaen}
\delta \bar{E}_{\N} (\mu_{1,2},\mu) \equiv \frac{E_{\N
max}(\mu_{1,2},\mu) -E_{\N min}(\mu_{1,2},\mu)}{E_{\N
max}(\mu_{1,2},\mu) +E_{\N min}(\mu_{1,2},\mu)}\,.
\end{equation}
It is easily seen that this error decreases exponentially both with
increasing global purity and decreasing marginal purities, {\em
i.e.~}with increasing entanglement. For ease of graphical display,
let us consider the important case of symmetric Gaussian states, for
which the reduction $\mu_1 = \mu_2 \equiv \mu_i$ occurs.
\begin{figure}[t!]
\begin{minipage}[t]{7cm}
\centering
 \includegraphics[width=7cm]{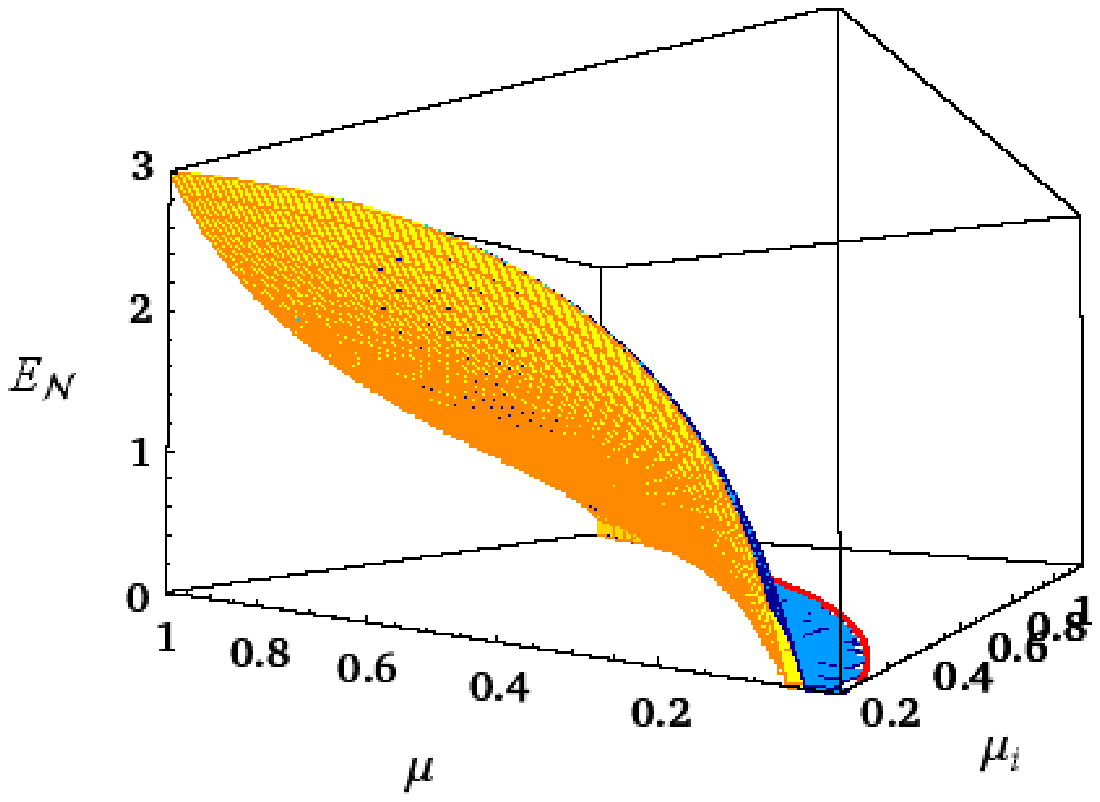}\\
  \caption{\footnotesize Maximal and minimal logarithmic
  negativity as functions of the global and marginal purities of
  symmetric Gaussian states. The darker (lighter) surface represents GMEMS
  (GLEMS).}
  \label{fig3D}
\end{minipage} \hspace*{.5cm}
\begin{minipage}[t]{6cm}
\centering
\includegraphics[width=6cm]{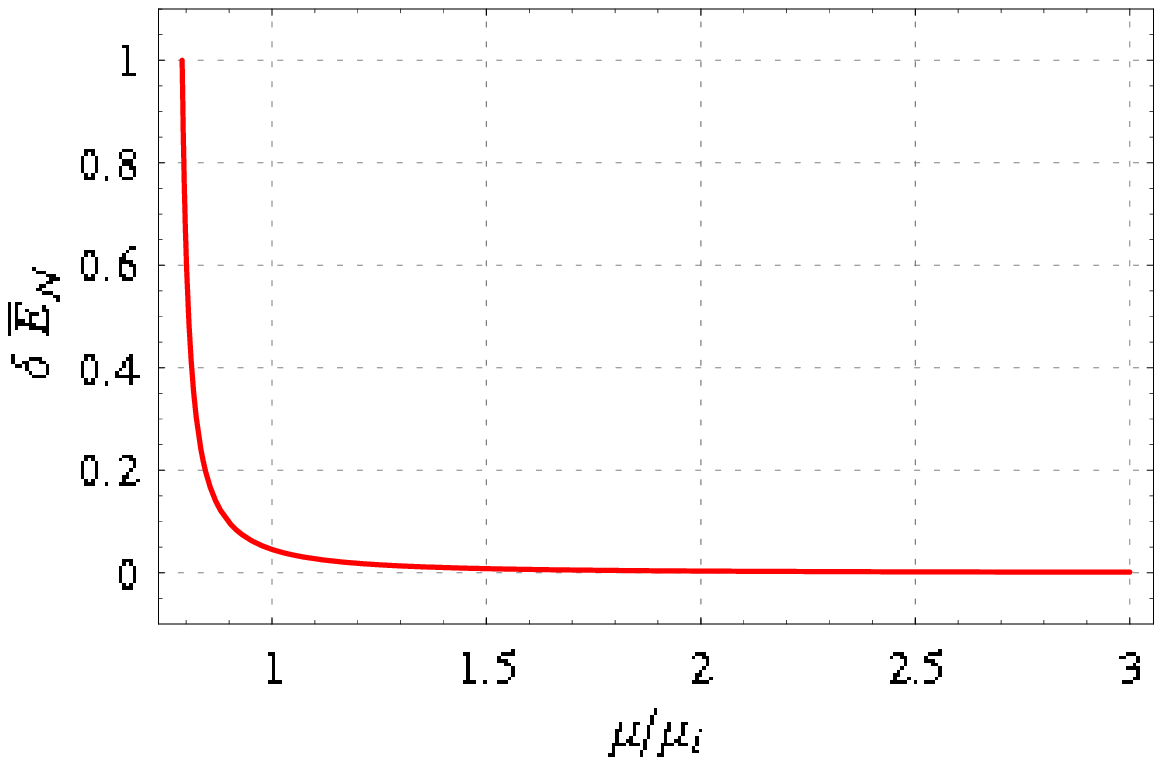}\\
  \caption{\footnotesize Plot of relative error $\delta \bar{E}_{\N}$
on the average logarithmic negativity as a function of the ratio
$\mu / \mu_i$ ($\mu=0.5$).}
  \label{figerror}
\end{minipage}
\end{figure}
 In Fig.~\ref{fig3D},
$E_{{\N}max}(\mu_i,\,\mu)$ of Eq.~(\ref{enmax}) and
$E_{{\N}min}(\mu_i,\,\mu)$ of Eq.~(\ref{enmin}) are plotted versus
$\mu_i$ and $\mu$. In the case $\mu=1$ the upper and lower bounds
correctly coincide, since for pure states the entanglement is
completely quantified by the marginal purity. For mixed states this
is not the case, but, as the plot shows, knowledge of the global and
marginal purities strictly bounds the entanglement both from above
and from below. The relative error $\delta \bar{E}_{\N}
(\mu_i,\,\mu)$ given by Eq.~(\ref{deltaen}) is plotted in
Fig.~\ref{figerror} as a function of the ratio $\mu / \mu_i$. It
decays exponentially, dropping below $5\%$ for $\mu > \mu_i$. Thus
the reliable quantification of quantum correlations in genuinely
entangled states is always assured by this method, except at most
for a small set of states with very weak entanglement (states with
$E_{\N} \lesssim 1$). Moreover, the accuracy is even greater in the
general non-symmetric case $\mu_1 \neq \mu_2$, because the maximal
entanglement decreases in such an instance (see Fig.~\ref{gmemms}).
This analysis shows that the average logarithmic negativity
$\bar{E}_{\N}$ is a reliable estimate of the logarithmic negativity
$E_{\N}$, improving as the entanglement increases. This allows for
an accurate quantification of CV entanglement by knowledge of the
global and marginal purities. The purities may be in turn directly
measured experimentally, without the full tomographic reconstruction
of the whole CM, by exploiting quantum networks techniques \cite{network}
or single--photon detections without homodyning \cite{cerf,wenger}.

\subsection{Entanglement vs Information (IV) --
Extremal entanglement at fixed global and local generalized
entropies}

In this section we introduce a more general characterization of the
entanglement of two--mode Gaussian states in terms of the degrees of
information, by exploiting the generalized Tsallis $p-$entropies
\begin{equation}\label{pgen}
S_{p} \equiv \frac{1-\,{\rm Tr}\,\varrho^p}{p-1} \; , \quad p > 1,
\end{equation}
as measures of global and marginal mixedness. Such an analysis can
be carried out along the same lines of the previous section, by studying
the explicit behavior of the global invariant $\Delta$ at fixed
global and marginal entropies, and its relation with the logarithmic
negativity $E_\N$. Let us remark that the $S_p$'s can be computed
for a generic Gaussian state in terms of the symplectic eigenvalues
\cite{asipra04}, namely
\begin{equation} {\rm
Tr}\,\varrho^{p}=\prod_{i=1}^{n}g_{p}(\nu_i)\; , \quad
g_{p}(x)=\frac{2^p}{(x+1)^p-(x-1)^p} \, . \label{pgau}
\end{equation}

We begin by observing that the standard form CM
$\gr\sigma$ of a generic two--mode Gaussian state \eq{stform} can be
parametrized by the following quantities: the two marginals
$\mu_{1,2}$ (or any other marginal $S_{p_{1,2}}$ because all the
local, single-mode entropies are equivalent for any value of the
integer $p$), the global $p-$entropy $S_p$ (for some chosen value of
the integer $p$), and the global symplectic invariant $\Delta$.

\begin{figure}[t!]
\centering \subfigure[\label{fig3D1}]
{\includegraphics[width=6.4cm]{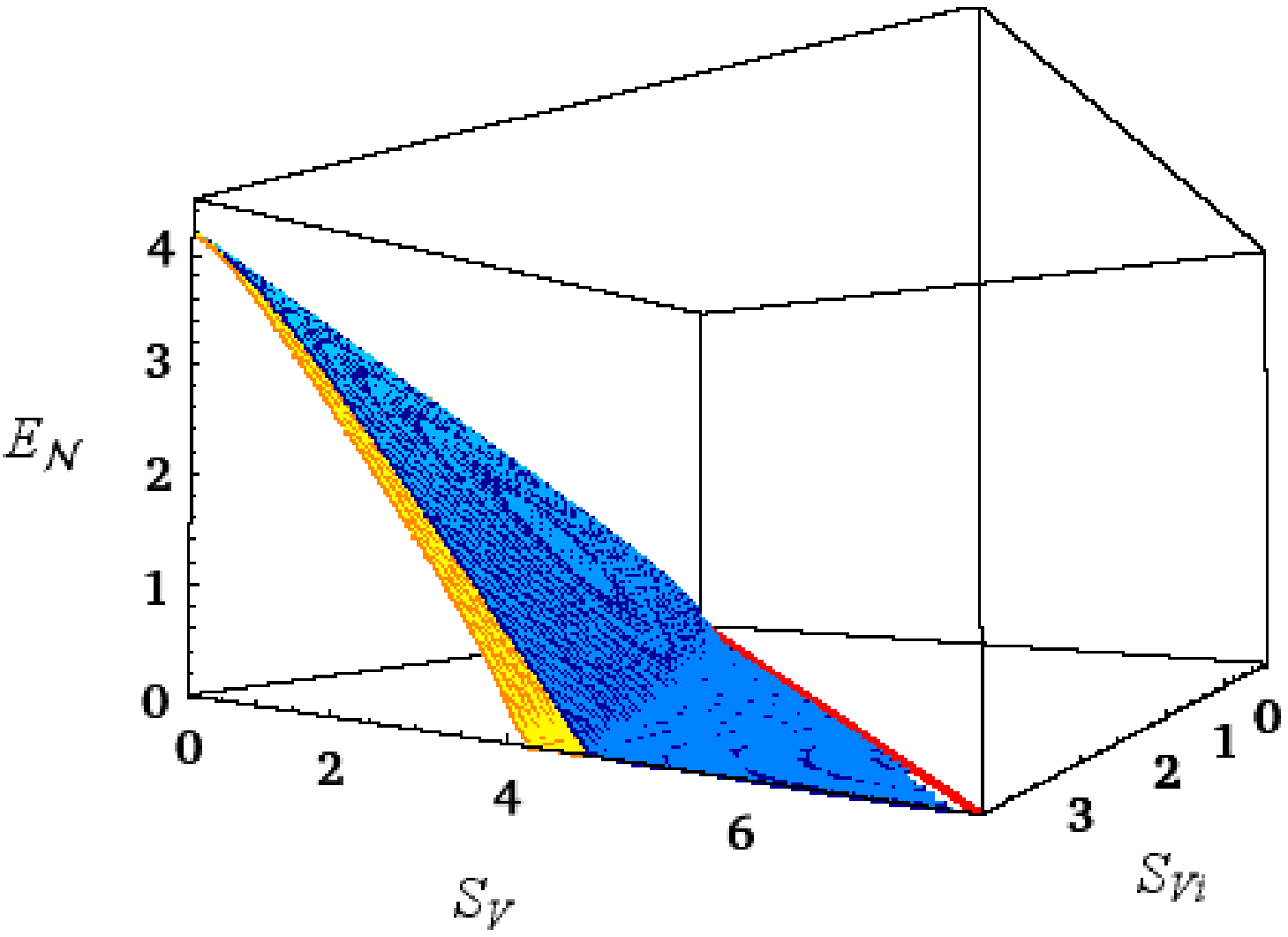}} \hspace{3mm}
\subfigure[\label{fig3D2}]
{\includegraphics[width=6.4cm]{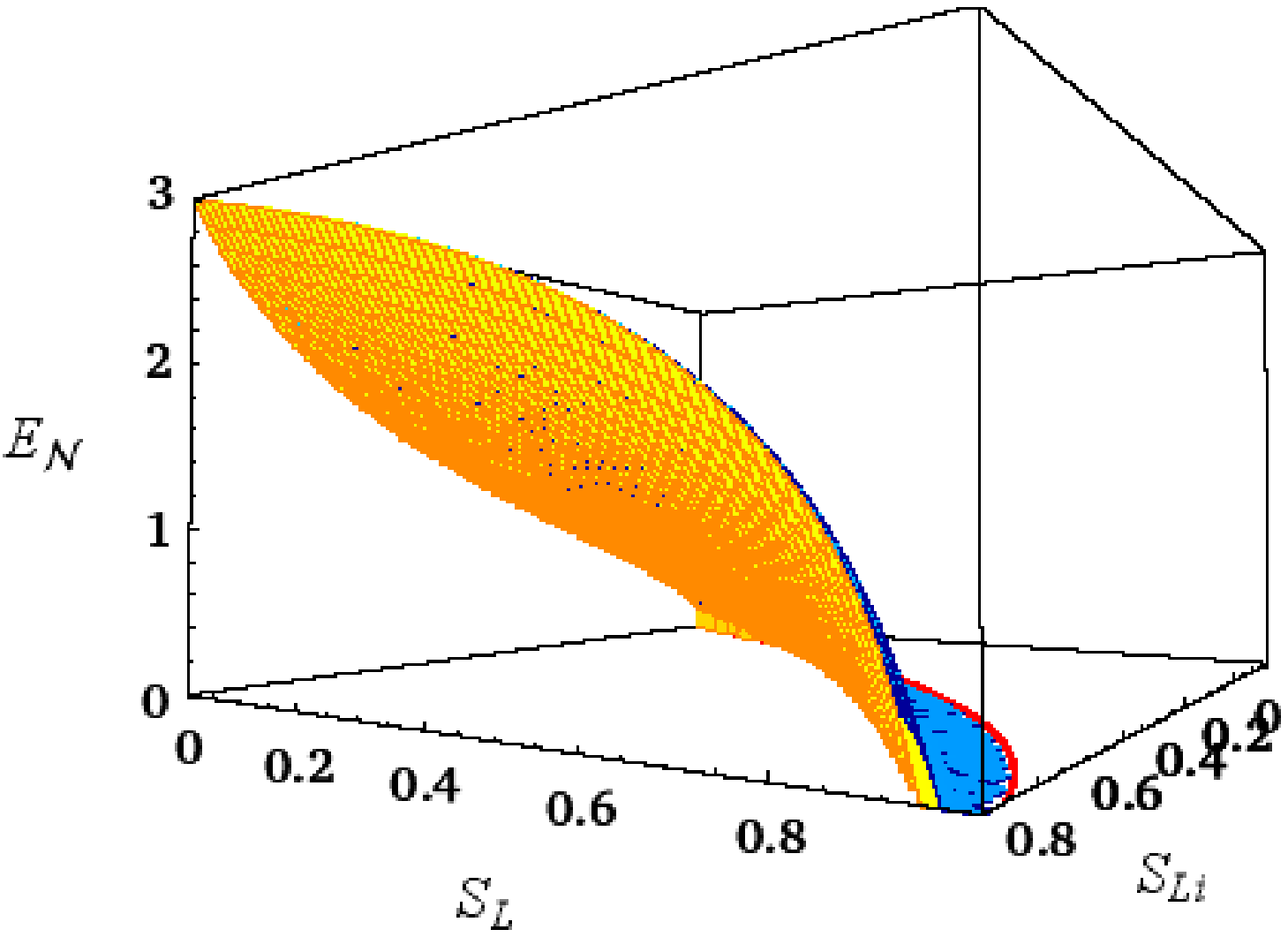}}
\subfigure[\label{fig3D3}]
{\includegraphics[width=6.4cm]{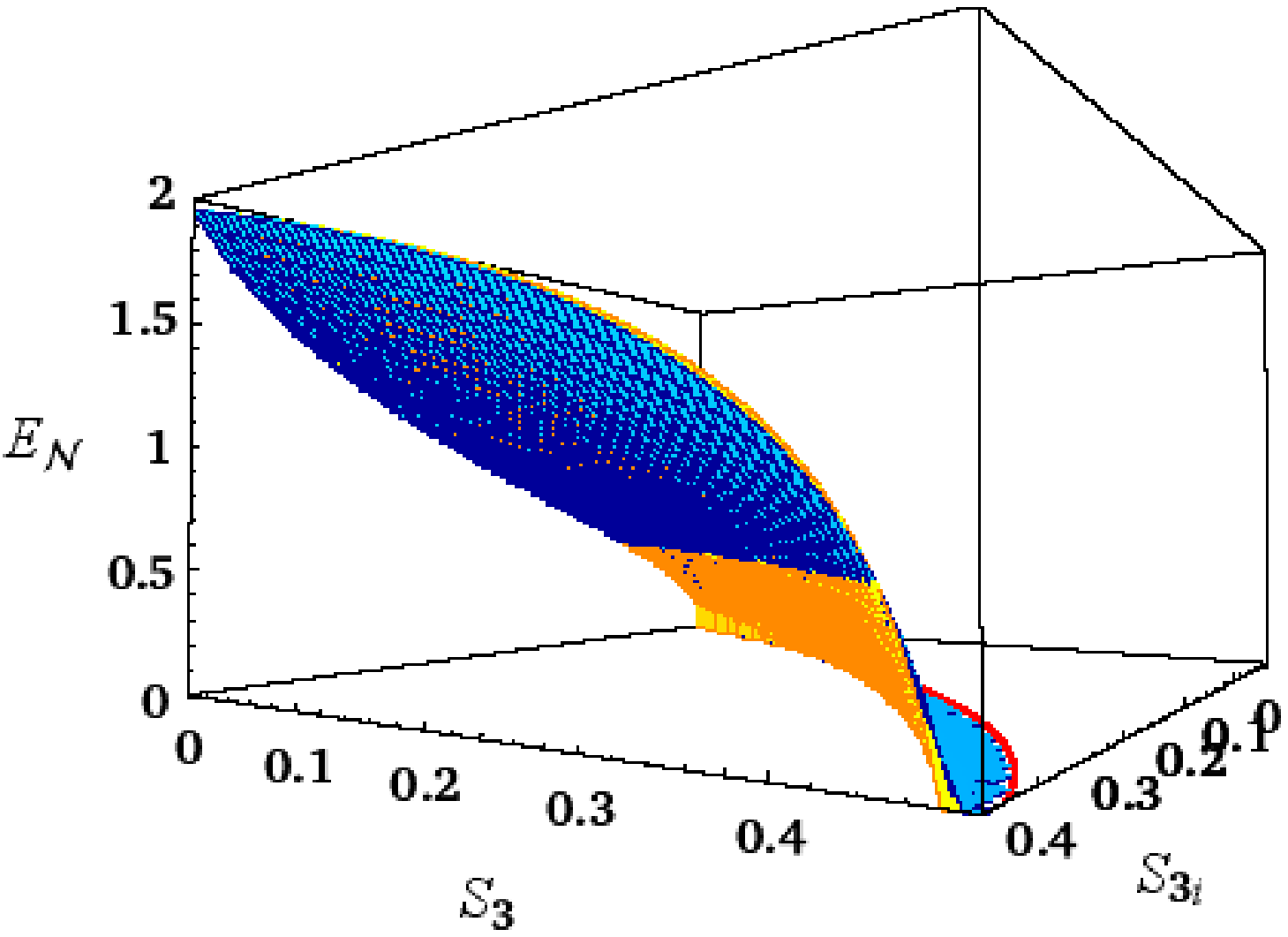}} \hspace{3mm}
\subfigure[\label{fig3D4}]
{\includegraphics[width=6.4cm]{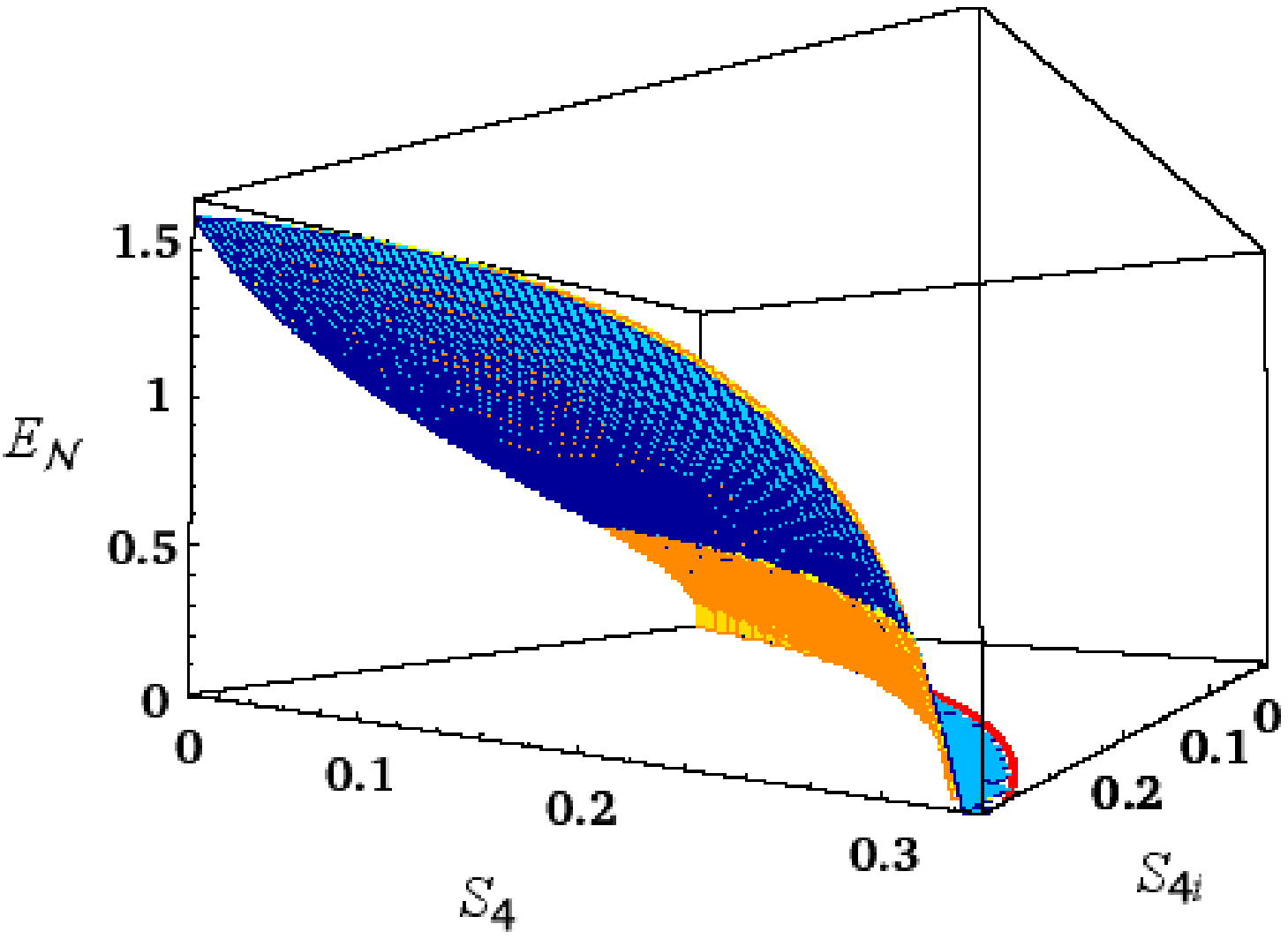}}
  \caption{\footnotesize Lower and upper bounds on the logarithmic
  negativity of symmetric Gaussian states
  as functions of the global and marginal generalized $p-$entropies, for
  (a) $p\rightarrow1$ (von Neumann entropies), (b) $p=2$ (linear entropies), (c) $p=3$,
  and (d) $p=4$. The darker (lighter) surfaces represent GMEMS
  (GLEMS). Notice that for $p>2$ GMEMS and GLEMS surfaces intersect
along the nodal inversion line (meaning they are equally entangled
along that line), and beyond it they interchange their role.}
  \label{fig3Dm}
\end{figure}

After somewhat lengthy but straightforward calculations (the details can be found in
Ref.~\cite{asipra04}), one finds that the entanglement is still bounded from above and from
    below by functionals of the global and marginal $p-$entropies,
    and the two extremal classes of states are again the nonsymmetric squeezed
    thermal states (GMEMS) and the mixed states of partial minimum uncertainty
    (GLEMS). Nevertheless, the seralian $\Delta$ is no
    longer monotonically related to the entanglement of the state, at fixed generalized
    entropies. In particular, for any $p>2$ ({\em i.e.~}with the exception of the
    linear and Von Neumann entropies), there exists a unique \emph{nodal
    surface} ${\cal S}_p^n(S_{p_{1,2}})$ such that
    \begin{equation}\label{cases}
\left.\frac{\partial\,\tilde\nu_-}{\partial\,
\Delta}\right|_{S_{p_1},S_{p_2},S_p} \quad {\rm is}
\quad\left\{\begin{array}{ll}
    >0, & \hbox{ when}\quad S_p>{\cal
S}_p^n(S_{p_{1,2}})\,; \\
    =0, & \hbox{ when}\quad S_p={\cal
S}_p^n(S_{p_{1,2}})\,; \\
    <0, & \hbox{ when}\quad S_p<{\cal
S}_p^n(S_{p_{1,2}})\,. \\
\end{array}
\right.
\end{equation}
While in the first case of \eq{cases} GMEMS and GLEMS retain their
property of being, respectively, maximally and minimally entangled
Gaussian states for fixed degrees of information, in the last case
they exchange their role: two--mode squeezed states become minimally
entangled states for fixed $p-$entropies, and the states of partial
minimum uncertainty are those with maximal entanglement. Even more
remarkably, the entanglement of all Gaussian states (including
GMEMS, GLEMS and infinitely many other different states) whose
$p-$entropies lay on the nodal surface of inversion $S_p={\cal
S}_p^n(S_{p_{1,2}})$, does not depend on $\Delta$ and is therefore
{\em completely} quantified in terms of the global and marginal
generalized degrees of information. Thus the state parametrization
by the Tsallis $p-$entropies (and, presumably, by the companion
family of the R\'enyi entropies \cite{zicoshan} as well), provides a
remarkable inversion between GMEMS and GLEMS, and allows to identify
a class of Gaussian states whose entanglement is quantified {\em
exactly} by the knowledge of the global and marginal degrees of
information. Moreover, even outside the nodal surface of inversion,
the gap between maximal and minimal entanglement for fixed
$p-$entropies decreases with increasing $p$, so that the accuracy of
the quantitative estimate of the logarithmic negativity as a
function of the $p-$entropies increases accordingly, as shown in
Fig.~\ref{fig3Dm}. Despite this interesting features, measurements
of purity, or equivalently of linear entropy ($p=2$), remain the
best candidates for a direct estimation of CV entanglement in
realistic experiments. This is due to the fact that measuring the
global $p-$entropy ($p\ne2$) of a state requires the full
tomographic reconstruction of the whole CM, thus nullifying the
advantages of the analysis presented in Sec.~\ref{secprl}.

\section{Quantification of multimode entanglement under symmetry}\label{secmulti}

For two--mode Gaussian states of CV systems we have shown that the measurement of the three
purities (or generalized entropies), out of the four independent
standard form covariances, suffices in providing a reliable quantitative
characterization of the entanglement. It is intuitively evident that the
efficiency of such a quantitative estimation in terms of
information entropies should improve significantly with increasing number
$N$ of modes, because the ratio between the
total number of covariances and the total number of global and
marginal degrees of information quickly scales with $N$. Moreover,
the structure of multipartite entanglement that can arise in multimode
settings is much richer than the basic bipartite CV entanglement.
Here we briefly discuss the simplest multipartite setting
in which the purities again successfully bound the entanglement
with great accuracy.

We consider highly symmetric $(N+1)$-mode Gaussian states of generic
single-mode systems $\gr\alpha$ coupled to fully symmetric $N$-mode
systems $\gr{\sigma}_{\beta^N}$, resulting in a global CM
$\gr\sigma$ of the form  \vspace*{-0.2cm}
\begin{equation}\label{sig}
\gr\sigma\!=\!\left(%
\!\begin{array}{cc}
  \gr\alpha & \gr\Gamma \\
  \gr\Gamma^{\sf T} & \gr\sigma_{\beta^N} \\
\end{array}%
\right)\!,\quad\gr\sigma_{\beta^N}=\left(%
 \begin{array}{cccc}
  \gr\beta & \gr\varepsilon & \cdots & \gr\varepsilon \\
  \gr\varepsilon & \gr\beta & \gr\varepsilon & \vdots \\
  \vdots & \gr\varepsilon & \ddots & \gr\varepsilon \\
  \gr\varepsilon & \cdots & \gr\varepsilon & \gr\beta \\
\end{array}%
\right)\!,\quad
\gr\Gamma=(\underbrace{\gr\gamma\;\cdots\;\gr\gamma}_N)\,.
\end{equation}
The state $\gr{\sigma}$ is determined by six independent parameters,
three of which are related to the fully symmetric $N$-mode block
$\gr{\sigma}_{\beta^N}$ (or, equivalently, to any of its two--mode
subblocks $\gr{\sigma}_{\beta^2}$). The remaining parameters are
determined by the single-mode purity $\mu_\alpha \equiv
(\det\gr\alpha)^{-1/2}$ and by the two global
$Sp_{(2N+2,\mathbbm{R})}$ invariants $\det\gr{\sigma} \equiv
1/(\mu_\sigma^2) = \prod_{i=1}^{N+1} \nu_i^2$ and $\Delta_\sigma
\equiv \Delta_{\alpha} + \Delta_{\beta^N} =\sum_{i=1}^{N+1}
\nu_i^2$. Here $\mu_\sigma$ is the global purity of the state
$\gr{\sigma}$, the $\nu_i$'s constitute its symplectic spectrum,
$\Delta_{\alpha} \equiv \det\gr\alpha+2N\det\gr\gamma$, and
$\Delta_{\beta^N} \equiv N (\det\gr\beta+(N-1)\det\gr\varepsilon)$.

To compute the multimode $1\times N$ logarithmic negativity of state
$\gr\sigma$ between
the mode $\gr\alpha$ and the $N$-mode block $\gr\sigma_{\beta^N}$,
it is convenient to perform the local symplectic operation that
brings $\gr\sigma_{\beta^N}$ in its Williamson normal form, which is
characterized by a $(N-1)$-times degenerate eigenvalue
$\nu_-=\sqrt{(b-e_1)(b-e_2)}$ (the same smallest symplectic
eigenvalue of the two--mode subblock $\gr\sigma_{\beta^2}$) and a
nondegenerate eigenvalue $\nu_{+^{(N)}} =
\sqrt{(b+(N-1)e_1)(b+(N-1)e_2)} = 1/(\nu_-^{N-1} \mu_{\beta^N})$,
where $\mu_{\beta^N}$ is the purity of the $N$-mode fully symmetric
block \cite{asimulti}. The crucial point here is that this local
operation actually decouples the mode $\gr\alpha$ from all the
$(N-1)$ modes corresponding to the eigenvalue $\nu_-$ and concentrates
the whole $1\times N$ entanglement between only two modes. The state
\eq{sig} is thus brought in the form
$\gr\sigma=\left(\begin{array}{cc}
  \gr\sigma^{eq} & {\bf 0} \\
  {\bf 0} & \gr\nu_- \\
\end{array}%
\right),$ where $\gr\nu_-$ is a diagonal $(2N-2)\times(2N-2)$ matrix
with all entries equal to $\nu_-$, and the \emph{equivalent}
two--mode state $\gr\sigma^{eq}$ is characterized by its invariants
\begin{equation}\label{inveq}
  \mu_1^{eq} = \mu_\alpha,\quad \mu_2^{eq} = \nu_-^{N-1}
  \mu_{\beta^N}, \quad \mu^{eq} = \nu_-^{N-1}\mu_\sigma, \quad
  \Delta^{eq}= \Delta_\alpha+(\nu_-^{N-1}\mu_{\beta^N})^{-2}\,.
\end{equation}
These invariants are, in turn, determined in terms of the six
invariants of the original $N$-mode state $\gr\sigma$, but we can
exploit the previous two--mode analysis (see Sec.~\ref{secprl}) to
conclude that, even without the explicit knowledge of $\Delta^{eq}$,
and so of the coupling $\det\gr\gamma$, the multimode entanglement
under symmetry can be quantified through the average logarithmic
negativity \eq{average} of the equivalent state $\gr\sigma^{eq}$,
with the \emph{same} accuracy demonstrated for generic two--mode
states (see Fig.~\ref{figerror}).

Thus the degrees of information of a multipartite Gaussian state
again provide a strong characterization and a reliable quantitative
estimate of the entanglement. Moreover, the method of the two--mode
reduction can be used to compute the $1\times K$ entanglement
between the mode $\gr\alpha$ and any $K-$mode subblock of
$\gr\sigma_{\beta^N}$, with $K=1,\ldots,N$, in order to establish a
multipartite entanglement hierarchy in the $(N+1)$-mode state of the
form \eq{sig} \cite{notehier}.

\section{Summary and Outlook}\label{secconcl}
In this work we aimed at unveiling the close relation between the
entanglement encoded in a quantum state and its degrees of
information. We have shown in detail how the knowledge of the global
degree of information alone, or of the marginal informations related
to the subsystems of a multipartite system, results in a qualitative
characterization of the entanglement: the latter increases with
decreasing global mixedness, with increasing marginal mixednesses,
and with marginal mixednesses as close as possible. We then proved
how the simultaneous knowledge of all the global and local degrees
of information of a Gaussian state leads to the identification of
extremally (maximally and minimally) entangled states at fixed
mixednesses (purities or generalized information entropies),
providing an accurate quantitative characterization of CV
entanglement. It is worth remarking that, out of the subset of
Gaussian states, very little is known about entanglement and
information in generic states of CV systems. Nevertheless, most of
the results presented here (including the sufficient conditions for
entanglement based on information measures), derived for CM's using
the symplectic formalism in phase space, retain their validity for
generic states of CV systems. For instance, any two-mode state with
a CM corresponding to an entangled Gaussian state is itself
entangled too \cite{vanlok}. So our methods may serve to detect
entanglement in a broader class of states of infinite-dimensional
Hilbert spaces.

The generalization of this analysis, connecting entanglement and
information, to highly symmetric multimode Gaussian states of $1
\times N$-mode partitions has been briefly sketched, presenting a
simple method to estimate CV multimode entanglement by measurements
of purity in an equivalent two--mode state. The extension of this
method to the quantification of multipartite entanglement in
Gaussian states with respect to generic $M \times N$ bipartitions of
the modes \cite{asimn}, as well as a deeper understanding of the
structure of genuine multipartite CV entanglement \cite{contangle}
and its relation with multiple degrees of information beyond the
symmetry constraints, are being currently investigated.

\section*{Acknowledgements}
\noindent G. A. acknowledges stimulating discussions with Frank
Verstraete, Reinhard F. Werner and Karol \.Zyczkowski at EIN04 in
Krzy\.zowa.


\begin{thebibliography}{99}
\bibitem{notepad} The largest eigenvalue $\tilde{\nu}_+$ of the
partially transposed CM of a two--mode Gaussian state is always
greater than 1 and so it does not enter in the computation of the
logarithmic negativity \eq{ensy}, see Ref.~\cite{asipra04}.
\bibitem{notehier} Being all equivalent to two--mode
entanglements, all the $1\times K$ entanglements can be directly
compared to each other. For furter details see Ref.~\cite{asimulti}.
\end{thebibliography}
\end{document}